\documentclass[aps,pre,twocolumn,groupedaddress,amsmath]{revtex4}

\usepackage[dvips]{graphicx,color}
\newcommand{\diff}{\mathrm{d}}

\begin{document}


\title{Folding and unfolding kinetics of a single semiflexible polymer}


\author{Natsuhiko Yoshinaga}
\affiliation{Department of Physics, Graduate School of Sciences, the
University of Tokyo, Tokyo 113-0033, Japan}


\date{\today}

\begin{abstract}
We theoretically investigate the kinetics of the folding transition of a single semiflexible
polymer.
In the folding transition, the growth rate decrease with an increase in the number of monomers
 in a collapsed domain, suggesting that the main contribution to
 dissipation is from the motion of the domain.
In the unfolding transition, dynamic scaling exponents, $1/8$ and $1/4$, were
determined for disentanglement and relaxation steps, respectively.
We performed Langevin dynamics simulations to test our theory.
It is found that our theory is in good agreement with simulations.
We also propose the kinetics of the transitions in the presence of the hydrodynamic interaction.
\end{abstract}


\maketitle


\section{Introduction}
Compared to our understanding of the equilibrium states of polymers, our
understanding of far from
equilibrium is quite primitive.
Recently, the kinetic properties of single polymer molecules have become
experimentally tractable, and they have been investigated mainly because
of their
biological importance.
Although the kinetics of ordering are well understood and several investigations
have been carried out on flexible polymers,
the kinetics of semiflexible polymers are not as well understood.

Most biopolymers, including DNA and many proteins, have bending
rigidity; therefore, they are classified as semiflexible polymers.
It is expected that bending rigidity plays an important role in
the structural stability and function of these biomacromolecules.
Recent experiments and simulations have clarified that a
single semiflexible polymer exhibits first-order phase transition between
an swollen coil state and a folded compact state as the solvent quality
decreases \cite{Takahashi:1997}.
A single semiflexible polymer folds into various kinds of ordered structures
depending on its bending rigidity and temperature, such as a
toroid or a cylinder \cite{noguchi:1996}.

The folding kinetics of a flexible polymer was first discussed by de Gennes
with the scaling theory \cite{degennes:1985,buguin:1996},
and then various methods have been proposed, such as Uniform Expansion Method
\cite{pitard:1998,pitard:1999} and Gaussian Self-consistent method 
\cite{timoshenko:1995}, where the latter was
extended to the folding kinetics of a semiflexible polymer \cite{kuznetsov:1996}.
Numerical simulations such as the Brownian Dynamics \cite{byrne:1995,chang:2001}
and Monte Carlo \cite{ostrovsky:1995,crooks:1999} simulations have been carried out.
Most of the simulations ignored the hydrodynamic
interaction. 
Recently, the authors in \cite{kikuchi:2002,kikuchi:2005} have
developed a new algorithm, which enables us to elucidate the features of the
hydrodynamic interaction in the kinetics of the collapse transition of a
flexible polymer.
These investigations have revealed that a flexible polymer forms a {\it
pearl-necklace} conformation at an early stage of a transition and
reaches the globule state via the growth of each pearl \cite{klushin:1998,halperin:2000}.
Although the globule is compact, it does not reach the most stable
state.
Therefore, at a later stage of the transition, segments realign so as
to form the equilibrium conformation \cite{grosberg:1988}.
The experiments with poly($N$-isoporpylacrylamide) (PNIPAM) were found to be consistent with the theoretical
and numerical results \cite{xu:2006}.

In this article, we discuss the transition kinetics
in single semiflexible polymers.
Our picture is that the folding transition consists of the nucleation and
growth steps, and the unfolding transition has three different
regimes: swelling, disentanglement, and relaxation \cite{yoshinaga:2006}.
Nucleation and growth processes have been
observed in the simulations of single semiflexible polymers and
experiments on single DNA molecules.
On the other hand, to our knowledge, the mechanism of the unfolding
transition has been much less elucidated, since it is difficult to observe the microscopic
dynamics of chain segments.
In \cite{rabin:1995,lee:2004}, authors theoretically and
computationnally proposed the existence of a topological constraint in single flexible polymers in the unfolding transition.
Our purpose is to study the time evolution of a macroscopic variable such as the long-axis length, which is experimentally observable.
We derive equation of motions for the variables and determine the dynamic scaling exponents in the unfolding
transition on the basis of a scaling analysis.
It is of importance to note that the validity of our picture can be verified by comparison
with exponents in experiments and simulations.

The reminder of this paper is organized as follows: In section
 \ref{section:theory}, we present an overview of our method. 
The results of the theoretical
 analysis of the folding and unfolding transitions are shown in section
 \ref{section:folding} and \ref{section:unfolding}, respectively.
Then, in section \ref{section:simulations}, we demonstrate the results of simulations and compare them with
 our theoretical results.
We discuss the hydrodynamic interaction in section
 \ref{section:hydrodynamics}.
Section \ref{section:pathway} is devoted to the justification of our theory
 with investigation of kinetic pathways.
In section \ref{section:summary}, we summarize our results. 

\section{Theory}
\label{section:theory}
In general, it is not trivial to write down the equations of motion of
course-grained variables such as a gyration radius.
In \cite{degennes:1985}, de Gennes proposed a powerful method to estimate
the time evolution of coarse-grained variables.
The essence of the method is based on the balance between the free
energy change of a polymer $F _{\rm chain}$ and the dissipative heat $Q$ due to the change.
The dissipative heat is exhausted by the motion of solvent
molecules; therefore, we have the following relation \cite{note1}
\begin{equation}
 \frac{{\rm d} F _{\rm chain}}{{\rm d} t}
= - \frac{{\rm d} Q }{{\rm d} t}.
\label{theory.balance}
\end{equation}
By calculating ${\rm d} F / {\rm d} t$ and ${\rm d} Q / {\rm d} t$
separately, we obtain the equation of motion for macroscopic variables \cite{degennes:1985,buguin:1996,halperin:2000}.

\textcolor[named]{Black}{
The dissipation arises from velocity gradient of fluid.
Nevertheless, if the hydrodynamic interaction is neglected (the so-called free-drain limit), the
dissipation is greatly simplified with Stokes force acting on solute molecules.
The Stokes force acting on a sphere of radius $b$ and a
velocity $v$ is $f \sim \eta v b$.
This leads to the dissipation
\begin{equation}
 \frac{{\rm d} Q _{\rm S}}{{\rm d} t}
\simeq \eta v^2 b.
\end{equation}
On the other hand, with hydrodynamic interactions, it is necessary to
consider velocity field.
Instead of solving full set of hydrodynamic equations,
we may consider two approximate situation;
When monomers in a domain moves cooperatively in one direction with the hydrodynamic interaction, the
frictional dissipation is still described by replacing size of a
particle with screening length $\xi_{\rm H}$.
We will further discuss this effect in section \ref{section:hydrodynamics}.
The second situation is a spherical expanding or contracting domain with
a size of $R_{\rm H}$, which has velocity gradient inside.
The force acting on a unit volume is proportional to gradient of
stress tensor, $\nabla \cdot
 \overleftrightarrow{\sigma}$, and this leads to the following
 dissipation from the domain with volume $\Omega$:
\begin{equation}
\frac{{\rm d} Q _{\rm H}}{{\rm d} t}
= \eta \int \diff \Omega (\nabla v)^2
\simeq \eta v^2 R_{\rm H}.
\label{eq:theory:viscos}
\end{equation}
}

\textcolor[named]{Black}{
We consider a polymer chain that has a contour length of $L$.
}
The total free energy has three contributions:
\begin{equation}
 F_{\rm chain} = F_{\rm ela} + F_{\rm bend} + F_{\rm int}.
\end{equation}
The first and second terms are entropic elasticity and bending
elasticity, respectively.
The third term arises from the repulsive interaction between monomers (the
excluded volume interaction) and, under a poor solvent condition,
attractive interaction.
The present theory does not successfully describe both the swollen and
the collapse states with unique approximate free energy.
Thus, we consider two asymptotic behaviors.
In the swollen state, we may combine the elastic and the bending free
energies and regard a renormalized monomer size as the persistence
length $l_p$ instead of the bare monomer size, $a$.
This implies that a polymer consists of the rods of length $l_p$.
Moreover, since the monomer density is low, $F_{\rm int}$ is expanded with
virial coefficients.
In the collapsed state, a mean field approximation is available \cite{grosberg:1992b}.
The total free energy has a contribution that is proportional to the volume
of a collapsed polymer.
With bending elasticity and surface penalty, the free energy is
described as 
\begin{equation}
 F_{\rm chain} = F_{\rm surface} + F_{\rm bending} + F_{\rm volume},
\end{equation}
where the surface energy is proportional to the surface area with a
surface tension $\sigma$.

In order to calculate free energy changes and dissipation, we assume
characteristic conformations in the folding and unfolding transitions.
In the folding transition, attractive interaction leads a polymer into a
folded state while all the other terms in the free energy prevent a polymer from making the transition.
This results in the competition between swelling and folding, and thus
we may expect the nucleation and growth steps in the folding
transition.
Contrary to this, no terms in the free energy stabilize the collapsed
state in the unfolding transition after switching the attractive
interaction off. 
Therefore, the instability at the early stage leads the unfolding transition.

The above assumption is justified by recent experiments on fluorescent
measurements of DNA, where  it has revealed that
the folding transition consists of the nucleation and growth steps,
while the kinetics
of the unfolding transition proceed more gradually \cite{yoshikawa:1996b}.
Simulations also support that the folding transition of a
single semiflexible polymer exhibits nucleation and growth \cite{sakaue:2002}.
Thus, it is reasonable to assume that the folding
transition consists of the nucleation and growth steps.
On the other hand, the details of the unfolding transition are yet unclear. 
We assume three regimes for the unfolding transition:
swelling, disentanglement, and relaxation steps.
Our purpose is to obtain the size of a polymer as function of time.
In the folding transition, we also calculate the nucleation time, which is
of importance to characterize the nucleation step.
These macroscopic values are relevant in comparison between methods
such as theory, simulations and experiments.
First, we will proceed our calculation under the assumption; then,
in section \ref{section:pathway}, we will discuss the validity of the assumption.

In this paper, we concentrate on the toroidal shape for $l_p \gg a$ and the globular shape
 for $l_p \approx a$, where $a$
is the monomer size, although,
in the folding transition of a semiflexible polymer, cylindrical
 conformations were also observed \cite{noguchi:1996}.
We make remarks concerning these conformations at the end of this paper.

\section{Folding Transition}
\label{section:folding}

\begin{figure}
\includegraphics{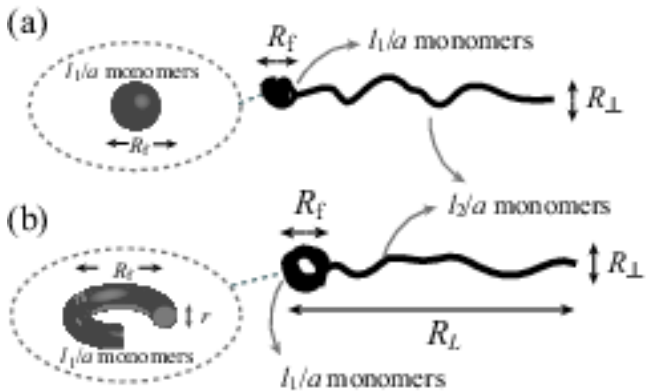}
\caption{Schematic representations of a polymer in the folding
 process at $l_p \approx a$ (a) and $l_p \gg a$ (b). The polymer consists of collapsed and coiled domains. The
 collapsed domain has $l_1/a$ monomers and the size $R_{\rm f}$. The coiled domain has
 $l_2/a$ monomers and the lateral fluctuation $R _{\perp}$.
\label{fig:theory_nucleation}
}
\end{figure}

The folding transition occurs as a result of the competition between the
volume free energy and the surface and the bending free energies.
The volume free energy makes a polymer collapse, while the surface and
the bending free energies lead a polymer to a coiled state.
Due to the competition, the folding transition is characterized with the nucleation and growth steps.
We consider a conformation shown in Fig. \ref{fig:theory_nucleation}.
When ${\rm d} l_1 / {\rm d} t > 0$, the nucleus grows and the system
reaches a collapsed state.
On the other hand, when ${\rm d} l_1 / {\rm d} t < 0$ the nucleus
becomes unstable and the system returns to a coiled state.
The free energy change and dissipation originates both from the
collapsed and coiled domains (Fig. \ref{fig:theory_nucleation}), 
\begin{eqnarray}
 \frac{{\rm d}F}{{\rm d}t} 
&=& 
\left( \frac{{\rm d}F}{{\rm d}t} \right)_{\rm collapse} 
+ \left( \frac{{\rm d}F}{{\rm d}t} \right)_{\rm coil},\\
 \frac{{\rm d}Q}{{\rm d}t}
&=&
\left( \frac{{\rm d}Q}{{\rm d}t} \right)_{\rm collapse}
+ \left( \frac{{\rm d}Q}{{\rm d}t} \right)_{\rm coil}.
\end{eqnarray}

\textcolor[named]{Black}{
We assume that the collapsed domain is close packed, {\it i.e}, its volume is
proportional to the length of the domain.
The conditions are $4/3 \pi R_f^3 = l_1 a^2$ for a flexible polymer and
$2 \pi ^2 r^2 R_{\rm f} = l_1 a^2$ for a semiflexible
polymer, where $R_f$ is the size of
the folded domain and $l_1/a$ monomers are inside the collapse (Fig. \ref{fig:theory_nucleation}).
A monomer at the boundary between collapsed and coiled domains
experiences the force
arising from the chemical potential of the collapsed domain,
\begin{equation}
 f = - \frac{\partial F_{\rm collapse}}{\partial l_1}.
\label{eq.fold.force}
\end{equation}
After nucleation, the monomers in the coiled domain are pulled into the
collapsed domain with this force.
}

\subsection{ \textcolor[named]{Black}{nucleation for a flexible chain ($l_p \approx a$)}}

When a polymer is flexible, {\it i.e.}, $l_p \approx a$,
it collapses into a disordered globule. 
Since, in the collapsed state, the mean field approximation is applicable,
the total free energy is written approximately with the contributions
from the surface, bending, and volume
free energies as 
\begin{eqnarray}
 F_{\rm collapse} &=& F_{\rm surface} + F_{\rm bend} + F_{\rm volume}\\
&\simeq& \sigma R_{\rm f}^2 + \kappa \frac{l_1}{R_{\rm f}^2} - \epsilon
 l_1 ,
\label{folding.flexible.freeenergy}
\end{eqnarray}
where $\sigma$ and $\epsilon$ are the surface tension and interaction
energy density, respectively, and $R_f \simeq (l_1 a^2)^{1/3}$.
We use the relation between the bending constant and the persistence length:
$\kappa / T = l_p $.
The free energy change is 
\begin{equation}
\frac{{\rm d} F}{{\rm d} t}
\simeq  
\left[ \frac{\sigma a^{4/3}}{l_1^{1/3}} + \frac{T l_p}{l_1^{2/3} a^{4/3}} - \epsilon \right] 
\left( \frac{{\rm d} l_1}{{\rm d} t} \right).
\label{eq_free_change_folding}
\end{equation}
Because the dissipation is an irreversible process, the sign of
${\rm d}l_1 / {\rm d}t$ is determined by the free energy change.
Therefore, the critical size of a nucleus is calculated as
\begin{equation}
 l_1 ^{*} \simeq 
\left( \frac{\sigma}{\epsilon} \right) ^3 a^4 
+ \frac{ 3 \sigma T l_p}{\epsilon^2}.
\end{equation}
The free energy at this size corresponds to the barrier between the
coiled and collapsed states, and the characteristic time for the
nucleation process $\tau _c$ exponentially depends on it.
In addition, $\tau _c$ weakly depends on the dissipation rate because
it determines the velocity of the growth of a nucleus. 
We obtain the nucleation time for a disordered collapse as 
\textcolor[named]{Black}{
\begin{equation}
 \tau _{c} 
\simeq 
\frac{\tilde{\gamma}^{*}}{\sigma} \left( \frac{l_1^{*}}{a} \right)^{4/3} 
\exp \left[ \frac{\sigma ^3 a^4}{\epsilon ^2 T}   
+ \frac{3 \sigma l_p}{\epsilon } \right],
\label{eq:fold:flexible:nucleation}
\end{equation}
where $\tilde{\gamma} ^{*} = \tilde{\gamma} (l_1 = l_1^{*})$ is the effective friction at the critical
nucleus.
The prefactor arises from second derivative of free energy at critical size of
nucleus.
}
The details of the prefactor depend on the dissipation mechanism which is
discussed below.
Nevertheless, the nucleation time is essentially dominated by the
exponential factor.

\subsection{nucleation for a semiflexible chain ($l_p \gg a$)}

The free energy of a collapsed domain depends on its structure and here we
consider a toroidal conformation (Fig. \ref{fig:theory_nucleation}).
The free energy is expressed as the summation of the surface,
bending and volume energies,
\begin{equation}
 F _{\rm collapse} ^{\rm toroid} 
\simeq \sigma r R_{\rm f} + \kappa \frac{l_1}{R_{\rm f}^2} - \epsilon l_1.
\end{equation}
The surface and
bending energies are balanced; as a result,
with the condition of the closely packed conformation in the folded part $2 \pi
^2 r^2 R_{\rm f} = l_1 a^2$, the free energy change is estimated as
\begin{equation}
 \left( \frac{{\rm d} F }{{\rm d} t} \right) _{\rm collapse} ^{\rm toroid}
\simeq   
\left[ \left( \frac{\sigma ^4 l_p T a^4}{l_1 ^2} \right) ^{1/5} - \epsilon \right] 
\left( \frac{{\rm d} l_1}{{\rm d} t} \right).
\label{folding.semiflexible.freeenergy}
\end{equation}

We obtain the following nucleation time for
ordered collapse:
\textcolor[named]{Black}{
\begin{equation}
\tau _{c}  
\simeq 
\frac{\tilde{\gamma}^{*} l_1^{*7/5}}{  (\sigma ^4 l_p T a^4)^{1/5}}  
e^{\epsilon l_1^*},
\label{folding.semiflexible.nucleationtime}
\end{equation}
}
where $l_1^* = \sigma ^2 l_p^{1/2} / \epsilon ^{5/2}$.

\subsection{growth}
In the growth step, the force $f$, given by (\ref{eq.fold.force}), that
pulls the monomers in the coiled domain is
balanced by the {\it effective} frictional force arising from dissipation:
\begin{equation}
 \tilde{\gamma} v = f,
\end{equation}
where $v=\diff l_1/\diff t = - \diff l_2/\diff t$.
The absorbing force of Eq.(\ref{eq.fold.force}), arising from free energy change, is essentially,
\begin{equation}
 f \simeq \epsilon,
\end{equation} 
where we neglect the contribution of the surface
term, which is sufficiently small at $l_1 \gg a$.
The effective friction depends on the dissipation mechanism, and
therefore, in general, would depend on the conformation of a chain.
The above force-balance relation is nothing but energy balance in
(\ref{theory.balance}) as can be confirmed by multiplying both sizes by $\diff l_1/\diff t$.

Let us consider the conformation shown in
Fig. \ref{fig:theory_nucleation} just after nucleation.
The collapsed domain is pulled by the coiled domain with $f$ and {\it
vice versa}.
The system is similar to the relaxation of stretched
polymers\cite{brochardwyart:1994,hallatschek:077804} and adsorption of
polymers by pulling on end with external force \cite{grosberg:228105,sakaue:2007}.
In the former system, flexible \cite{brochardwyart:1994} and semiflexible \cite{hallatschek:077804}
polymers are stretched when a force is
applied at one or two ends.
In the latter systems, a polymer is pulled at a constant force and is
absorbed into a pore.
In both systems, after the force is switched on, tension propagates along a chain with
finite velocity.
Therefore, not all the monomers move by the force, but
some parts are driven by the force and others do not feel it.
The results of these works show that the effective friction depends on
the length of monomers moving under force, and their motion is initiated
by the propagation of the force.

In our system, the force is acting on the interface between collapsed and
coil domains.
While the force quickly propagates on the entire collapsed domain due to
the cooperative motion of monomers, the force acting on the coil domain
propagates with some velocity.
The time required for propagation is not infinitesimal.
The overall motion of a chain does not contribute to the effective
friction, but relative motion of a collapsed or coil domain leads to
dissipation.
Thus, the question arises in the growth step: which domain moves?
This is equivalent to asking which domain contributes to dissipation.
Since in the free-drain limit, friction is proportional to the length of a
moving domain, the collapsed domain moves quickly and makes a dominant
contribution, at least, at an early stage
of the growth step. 
At a later stage, when the length of a collapsed domain becomes much longer
than that of a coil domain, and {\it in addition}, when the force propagates to the
free end of the coil domain, the coiled domain makes dominant
contribution to dissipation.
\textcolor[named]{Black}{
Therefore, we assume that the dominant contribution is from the
collapsed domain at the early stage.
The effective frictional coefficient  is described in the free-drain limit as
\begin{equation}
 \tilde{\gamma} \simeq \gamma_1 = \eta l_1.
\end{equation}
In semiflexible polymers, the logarithmic correction $(\ln(l_p/a))^{-1}$
is multiplied on the right-hand side.
At the later stage after force propagation, the effective friction is
described as 
\begin{equation}
 \tilde{\gamma} \simeq \gamma_2 = \eta ( L - l_1).
\end{equation}
Therefore, the velocity is 
\begin{equation}
v \simeq \left( \frac{1}{\gamma_1} + \frac{1}{\gamma_2} \right) f
= \frac{\epsilon L}{\eta l_1 (L - l_1)},
\label{eq.th.growth.l1}
\end{equation}
}
where we neglect the logarithmic correction.
Note that this argument is justified even with hydrodynamic
interaction. 
We will see this in section~\ref{section:hydrodynamics}.

\section{Unfolding Transition}
\label{section:unfolding}
\begin{figure}
\includegraphics{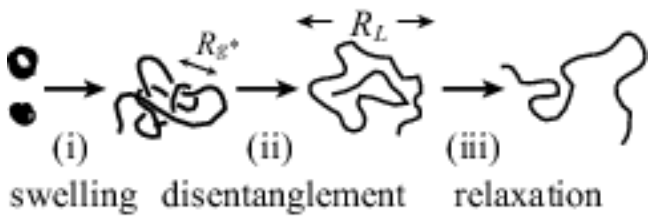}
\caption{Three steps in the unfolding process. (i) swelling, (ii)
 disentanglement, and (iii) relaxation. We characterize the time
 evolution with the size $R_L$.
\label{fig:theory_unfold}
}
\end{figure}

When, at $t=0$, the attractive interaction is switched off, a collapsed
polymer starts to unfold.
While, in the folding transition, the kinetics is dominated by the competition between the free
energies that do and do not prefer the collapsed state, in the unfolding transition
{\it all} the terms in the free
energy make a collapsed polymer unfold.
In fact, the free energies of the entropic elasticity, the bending elasticity, and the
excluded volume interaction are larger in the folded state.
This suggests that the collapsed state is unstable at an
early stage of the transition.
After the swelling step, the structure is similar to a coiled state, but it has many
entanglements, which are expected to lead to slow kinetics. 
Finally, a polymer relaxes into an equilibrium coiled state.
Therefore, we assume that the unfolding process consists of three steps: swelling,
disentanglement, and relaxation (fig. \ref{fig:theory_unfold}).

In the unfolding transition, the free energy is described with elastic
and interaction terms.
\begin{equation}
 F \simeq T(\alpha ^2 + \alpha ^{-2})
+ \frac{B}{R_L^3} \left( \frac{L}{l_p} \right)^2,
\label{unfolding.freeenergy}
\end{equation}
where $\alpha = R_L/(L l_p)^{1/2}$, and $B \simeq l_p a^2 T(1-\Theta/T)$ is the second virial coefficient.
$\Theta$ is the critical temperature of the transition.
At $t=0$, we increase the temperature so that $B \simeq l_p a^2 T >0$.
In the entropic contribution, the first term shows energy penalty for expansion and the second term
corresponds to that for compression.
Here we consider relaxation to an equilibrium coiled state, {\it i.e.},
the second term is dominant.
The volume interaction is dominant at the swelling step and
almost disappears after the step. 
In the disentanglement and relaxation steps, we consider the free energy
change due to the entropic contribution in Eq. (\ref{unfolding.freeenergy}).

\textcolor[named]{Black}{
\subsection{Swelling}
}

In this step, the dominant contribution of interactions between monomers
gives the free energy change,
\begin{equation}
 \frac{{\rm d} F }{{\rm d} t}
\simeq 
\left( - \frac{B L^2}{R_L^4 l_p^2} \right)
\frac{{\rm d} R_L}{{\rm d} t}.
\label{swelling.folding.freeenergychange}
\end{equation}
The dissipation arises from all the monomers and is described as,
\begin{equation}
\frac{{\rm d} Q _{\rm S}}{{\rm d} t}
\simeq \eta \frac{L}{l_p} \frac{l_p}{\ln (l_p/a)} 
\left( \frac{{\rm d} R_L}{{\rm d} t} \right) ^2.
\label{swelling.entropy}
\end{equation}
With Eq. (\ref{theory.balance}), the time evolution is given by
\begin{equation}
\frac{R_L}{R_{\rm swell}} \simeq 
\left( 1 + \frac{|B|L}{l_p^2 R_{\rm swell}^5}t \right)^{1/5},
\end{equation}
where $R_{\rm swell}$ is the size in folded states.

Since monomer-monomer interaction is short ranged (the length scale is $\delta
=\mathcal{O}(a)$), this regime is over when the mean distance between
monomers exceeds $\delta$.
Therefore, the characteristic time is 
\begin{equation}
 \tau _{\rm swell} 
\simeq 
 \frac{L^{2/3} l_p}{\ln(l_p/a) a^{5/3}} \tau_0,
\end{equation}
where we use the microscopic time scale $\tau_0 = \eta a^3/k_BT$.
As we will see later, this time scale ($\sim L^{2/3}$) is much shorter than the time
scale of other two regimes.
We conclude that the swelling regime does not exhibit dynamic scaling
behaviors.

\subsection{Disentanglement}
In the steps of disentanglement and relaxation, the free energy change originates from the elastic
free energy, which is given by
\begin{equation}
 \frac{{\rm d} F _{\rm ela}}{{\rm d} t} 
\simeq - \frac{T}{\alpha ^3}  \frac{{\rm d} \alpha}{{\rm d} t}.
\label{eq.free.change2}
\end{equation}

For the conformation with entanglements, a topological blob, which
consist of $g^* l_p$ monomers and
is $R_{g^*}$ in size can be defined \cite{grosberg:1988}.
Inside a topological blob, a chain behaves like a Gaussian coil, whereas on a
larger scale, a chain is assumed as a swollen globule.
Therefore, $R _{g^*} ^2  \sim g^* l_p ^2$ and $g^* l_p/R _{g^*} ^3  \sim L/R_L^3$ is satisfied.
This leads to the following relations:
\begin{equation}
 g^* 
\simeq \frac{R_L^6}{L^2 l_p ^4},
\end{equation}
\begin{equation}
 R_{g^*} \simeq \left( \frac{l_p g^*}{L} \right) ^{1/3} R_L .\label{rg_to_r}
\end{equation}
Beyond the scale, blobs are frozen due to entanglements and behave as obstacles when monomers in a particular blob are driven.
Since we neglect the hydrodynamic interactions, the dissipation inside
a blob is proportional to the number of segments $g^*$, and it is written
in the free-drain limit as
\begin{equation}
 \frac{{\rm d} Q _{\rm S}}{{\rm d} t}
\simeq 
\frac{L^2}{l_p^2 g^*} \eta g^* \frac{l_p}{\ln (l_p/a)} 
\left( \frac{{\rm d} R_{g^*}}{{\rm d} t} \right) ^2
\simeq
\eta \frac{R_L ^4}{l_p ^3 \ln (l_p/a)} 
\left( \frac{{\rm d} R_L}{{\rm d} t} \right) ^2.
\label{eq:fold:Qdisentangle}
\end{equation}
From the energy balance, we obtain
\textcolor[named]{Black}{
\begin{eqnarray}
 \frac{R_L}{R_{\rm dis}} 
&\simeq&
\left( \frac{t}{\tau _{\rm dis}} + 1 \right) ^{1/8}, \label{eq.theory.disentanglement.size}\\
\tau _{\rm dis}
&\simeq& 
 \frac{ L ^{5/3} l_p ^{4/3} }{a^{3} \ln (l_p/a)} \tau_0
\label{eq.theory.disentanglement.tau},
\end{eqnarray} 
where $R_{\rm dis} \simeq L^{1/3} l_p^{2/3}$ are the initial value and $\tau
_{\rm dis}$ is the
characteristic time scale, respectively, in this regime.
}
We have assumed the swollen globular structure for the initial state.
For $t \gg \tau _{\rm dis}$, the following scaling relation is obtained;
\begin{equation}
 R_L \sim t  ^{1/8}.
\label{eq.unfolding.disentanglement.scaling}
\end{equation}
This step proceeds during $g^{*} \ll L/l_p$, where a polymer has many
entanglements so that we assume the outside of blobs to be frozen.
The characteristic time scale of this step is obtained with the
condition $g^{*} = L/l_p$ and is proportional to
$L^3$, which is much longer than the time scales of the other two steps.

\begin{figure}
\includegraphics{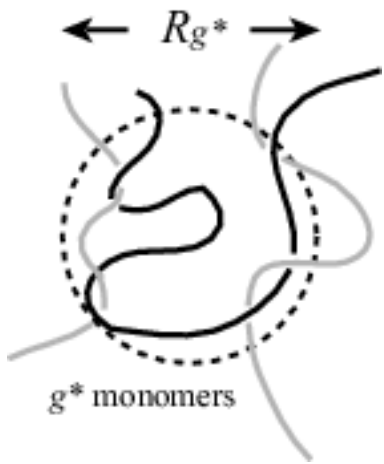}
\caption{Schematic representation of a topological blob of $g^*$
 monomers. Inside a blob, a polymer behaves as a coil, whereas it is
 entangled on a larger scale.
\label{fig:theory_disentangle}
}
\end{figure}

\subsection{Relaxation}
Contrary to the disentanglement step, in the relaxation
step, all segments contribute to dissipation, and thus, it is given as
\begin{equation}
 \frac{{\rm d} Q _{\rm S}}{{\rm d} t}
\simeq \eta \frac{L}{l_p} \frac{l_p}{\ln (l_p/a)} \left( \frac{{\rm d} R_L}{{\rm d} t} \right) ^2.
\label{eq.dissipation.relaxation}
\end{equation}
From Eqs. (\ref{eq.free.change2}) and
(\ref{eq.dissipation.relaxation}), we obtain
\begin{eqnarray}
  \frac{R_L}{R_{\rm relax}} 
&\simeq&
\left( \frac{t}{\tau _{\rm relax} } + 1 \right) ^{1/4},
\label{eq.theory.relaxation.size} \\
\tau _{\rm relax}
&\simeq&
\frac{ L ^{2} l_p}{a^3 \ln (l_p/a)} \tau_0,
\label{eq.theory.relaxation.tau}
\end{eqnarray}
where $R_{\rm relax} \sim (L l_p)^{1/2} $ and $\tau _{\rm relax}$ are the initial value and the characteristic time scale,
respectively, in the relaxation step.
We estimate $R_{\rm relax}$ from the condition
$g^* = L/l_p$.
We should note that the feature of $\tau _{\rm relax} \sim L^2$ is typical in the
Rouse dynamics \cite{doi:1986}.
For $t \gg \tau _{\rm relax}$, the following scaling relation is obtained:
\begin{equation}
 R_L \sim  t  ^{1/4}.
\label{eq.unfolding.swelling.scaling}
\end{equation}

\section{Simulations}   
\label{section:simulations}
In order to examine the folding and unfolding kinetics, we carried out Langevin dynamics
simulations of a bead-spring model using the following potentials:
\begin{eqnarray} 
V_{\mathrm{beads}} &=& \frac{k _a}{2} \sum _{i} ({| \mathbf{r} _{i+1} -
 \mathbf{r} _{i}|} - a)^2, \label{V_beads}\\
V_{\mathrm{bend}} &=& \frac{\kappa}{2} \sum _{i}  (1 - \cos \theta _{i} ),\\
V_{\mathrm{LJ}} &=& 4 \epsilon _{\mathrm{LJ}} \sum _{i,j} 
\left[ \left( \frac{a}{|\mathbf{r} _i - \mathbf{r} _j|} \right) ^{12} 
- \left( \frac{a}{|\mathbf{r} _i - \mathbf{r} _j|} \right) ^{6} \right],
\label{V_lj}
\end{eqnarray} 
where $V = V_{\mathrm{beads}} + V_{\mathrm{bend}} + V_{\mathrm{LJ}}$,
and $\mathbf{r} _{i}$ is the coordinate of the $i$th monomer, and $\theta
_{i}$ is the angle between adjacent bond vectors.
The monomer size $a$ and $k_{B} T$ are chosen as the unit length and energy,
respectively. 
Monomer-monomer interactions are included with the Lennard-Jones
potential, $V_{\mathrm{LJ}}$, which contains the soft-core excluded volume
interaction and the short-ranged attractive interaction. 
With small $\epsilon _{\mathrm{LJ}}$, the attraction is week such that
only the excluded volume interaction is relevant.
We set $\epsilon _{\mathrm{LJ}} = 0.3$ for this good solvent condition.
On the other hand, with large $\epsilon _{\mathrm{LJ}}$, the attractive
interaction plays a role for the folding transition (poor solvent condition).
We adopt the spring constant in $V_{\mathrm{beads}}$ to be $k_a = 400$.
The persistence length $l_p$
is a convenient measure to characterize the stiffness of a polymer
chain.
The bending elasticity $\kappa$ in $V_{\mathrm{bend}}$ satisfies the relation $\kappa = l_p T$.
We consider a homopolymer mainly with a polymerization index $N = 256$, which
has sufficient length for the formation of ordered structures (a toroid, a
cylinder, and so on) in a semiflexible polymer \cite{noguchi:1996}.

The equation of motion is written as
\begin{equation}
m \frac{{\mathrm{d}}^2 {\mathbf{r}} _i}{{\mathrm{d}} t^2} 
= - \gamma \frac{{\mathrm{d}} {\mathbf{r}} _i}{{\mathrm{d}} t}
- \frac{\partial V }{\partial {\mathbf{r}} _i}
+ {\boldsymbol{\xi}} _i,
\label{simulation}
\end{equation}
where $m$ and $\gamma $ are the mass and friction constant of monomeric units, respectively.
The unit time scale is $\tau_{0}^{s} = \gamma a^2/k_{B}T = 6 \pi \tau_{0}$.
We set the time step as $0.01 \tau_{0}^{s}$ and use $m=1.0$ and $\gamma =1.0$.
With these parameters, the relaxation time of the momentum of a monomer is
sufficiently fast as compared to the time scale of interest.
Gaussian white noise ${\boldsymbol{\xi _{i}}}$ satisfies the following
fluctuation-dissipation relation:
\begin{equation}
 <{\boldsymbol{\xi}} _i (t) \cdot {\boldsymbol{\xi}} _j (t')> = 6 \gamma
  k_{B}T \delta _{ij} \delta (t-t').
\end{equation}

The folding and unfolding states of a polymer are characterized by the
number of folded monomers $N_{\rm{fold}}$,
which is defined by
\begin{eqnarray}
\rho _{i} &=&  
\sum _j H ( r_c ^2 - | {\mathbf{r}}_i - {\mathbf{r}}_j |^2),\\
N_{\rm{fold}} &=& \sum _i \Theta (\rho _i - \rho _c),
\end{eqnarray}
where $H (x)$ and $\Theta (x)$ are the Heaviside and Step functions, respectively, and
$\rho _i$ is the local monomer density.
In this work, we set $r_c = 3.0$ and $\rho _c = 25.0$.
The nucleation time, $\tau _c$, is defined so as to satisfy
$N_{\rm{fold}} (\tau) \geq N_c$ for $\tau \geq \tau _{c}$,
where we set $N_c = 0.20$.
We should note that our results do not depend on these specific values.
Conformations of a polymer are characterized with the long-axis length,
\begin{equation}
 R_L = \max |{\bf r}_{i} - {\bf r}_{j}|.
\end{equation}

In the folding transition, coiled polymers were equilibrated under the good
solvent condition, $\epsilon _{\mathrm{LJ}} = 0.30$, and then quenched at $t=0$ into
the poor solvent conditions such as $\epsilon _{\mathrm{LJ}} = 0.70$, $1.0$, and
$1.3$. 
Although we had various structures such as a toroid and a rod for $l_p \geq 10a$,
 we chose toroidal conformations, and the macroscopic variables
were averaged over the ensemble of this conformation for consistency with
our theory.
To achieve this, we neglected the trajectories whose final conformations have the long-axis
length larger than $20a$ in order to ensure that the final conformations
are the troidal state.
Typically, cylindrical conformations have a size of more than $20 a$
in the folded state.
In the unfolding transition, we prepared folded polymers at $\epsilon
_{\mathrm{LJ}} = 1.0$, and quenched the system at $t=0$ into $\epsilon _{\mathrm{LJ}} = 0.30$.
We calculated the fraction of collapsed part and averaged it over more than 100 runs.

\begin{figure}
\includegraphics{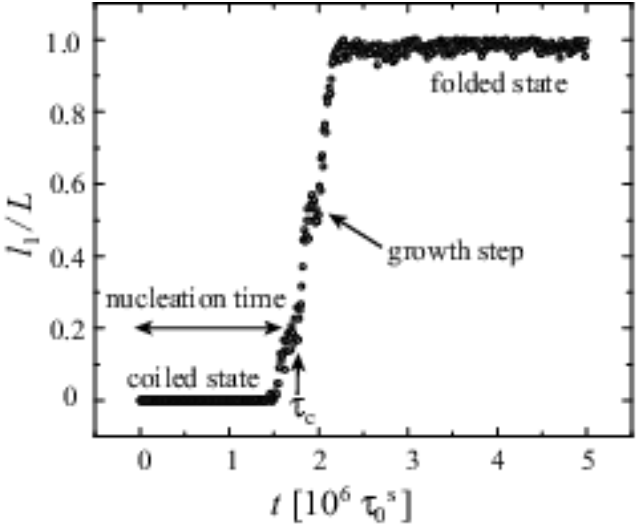}
\caption{
Typical time evolution of $l_1$ in the folding transition. The
 attractive interactions switch on at $t = 0 $ by replacing $\epsilon =
 0.3$ with $\epsilon = 1.0$.
\label{fig:simu_fold}
}
\end{figure}

\begin{figure}
\includegraphics{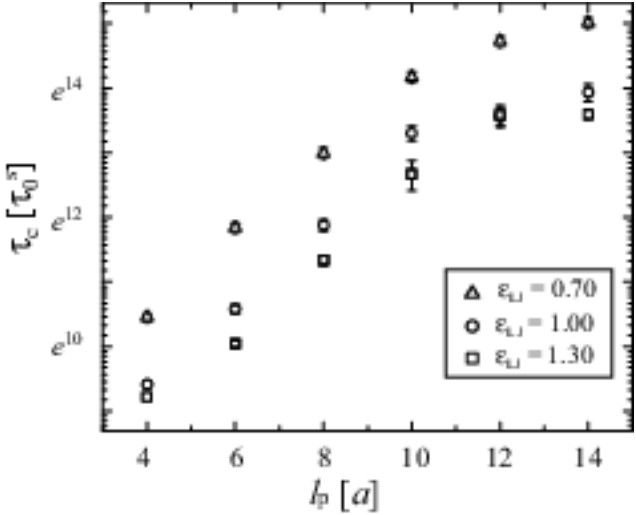}
\textcolor[named]{Black}{
\caption{A semi-log plot of nucleation time as a function of persistence length.
\label{fig:simu_nucleation}
}
}
\end{figure}

Figure \ref{fig:simu_fold} shows the typical time evolution of the
ratio of monomers in the folded state.
As we can see, after a long lag time, the ratio increases with
time and reaches the equilibrium value.
The dependence of the nucleation time on the persistence length is shown
in Fig. \ref{fig:simu_nucleation}.
For small $l_p$, the nucleation time exponentially depends on the
persistence length, which is consistent with Eq. (\ref{eq:fold:flexible:nucleation}). 
For large $l_p$, the slope becomes rather gentle. 
This is also consistent with our theory of
Eq. (\ref{folding.semiflexible.nucleationtime}), where $\tau _c \sim \exp (l_p^{1/2})$.

\begin{figure}
\includegraphics{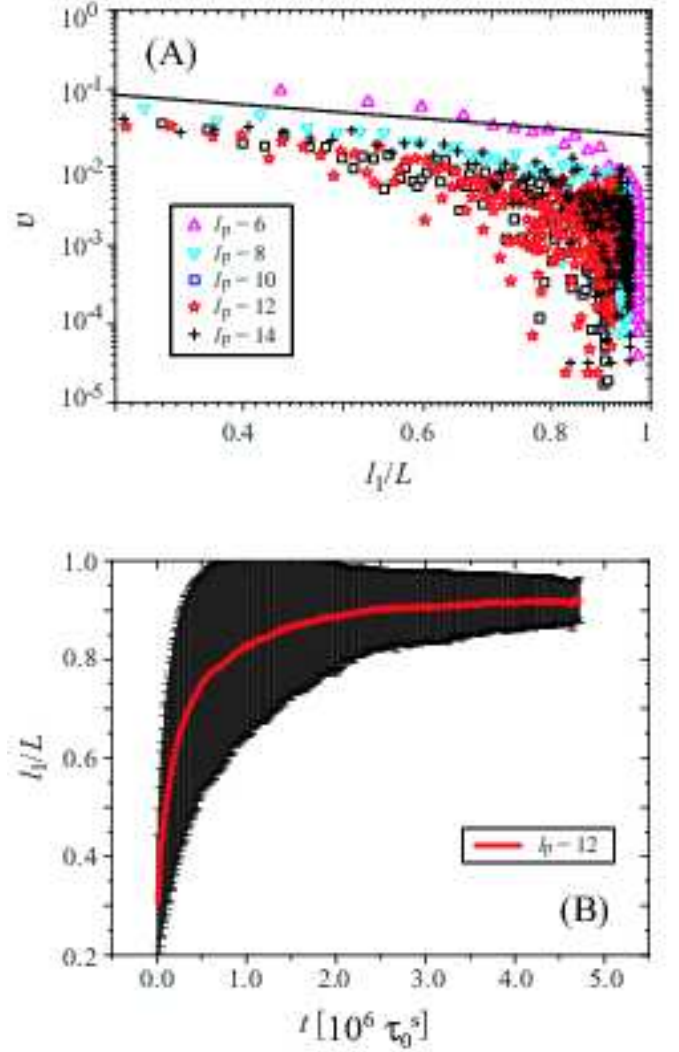}
\caption{\textcolor[named]{Black}{
(A) A log-log plot of velocity during growth steps. The line shows $v
 \sim l_1^{-1}$. (B) Mean value and fluctuation of the fraction of the folded state
 with $l_p = 12$.
}
\label{fig.simu.growth}
}
\end{figure}

The velocity of $l_1$ after nucleation is plotted in
fig. \ref{fig.simu.growth}, where the velocity is inversely proportional
to $l_1$.
This implies that friction does not arise from the coil part.
If the coil domain involves friction, the velocity will {\it increases}
with $l_1$.
\textcolor[named]{Black}{ 
In the section III C, we discussed velocity of the growth step, which is
valid early and later stages.
At the early stage, $v \sim \epsilon/(\eta l_1)$ while at the later
stage, $v \sim \epsilon/(\eta (L - l_1))$.
Our results of simulations are consistent with (\ref{eq.th.growth.l1})
at the early stage.
However, at the later stage, our theory predict increase of velocity. 
The discrepancy may be explained with large fluctuation of the fraction
of folded monomers.
In fact, the system
reaches at the folded state even at $l_1/L < 1$ as shown in Fig. \ref{fig.simu.growth}B.
Since the driving force for the folding transition decreases close to the
equilibrium state, velocity near $l_1/L \approx 1$ is expected to
decrease as in Fig. \ref{fig.simu.growth}.
In our theory, we assume that driving force is constant throughout the
growth step.
}

\begin{figure}
\includegraphics{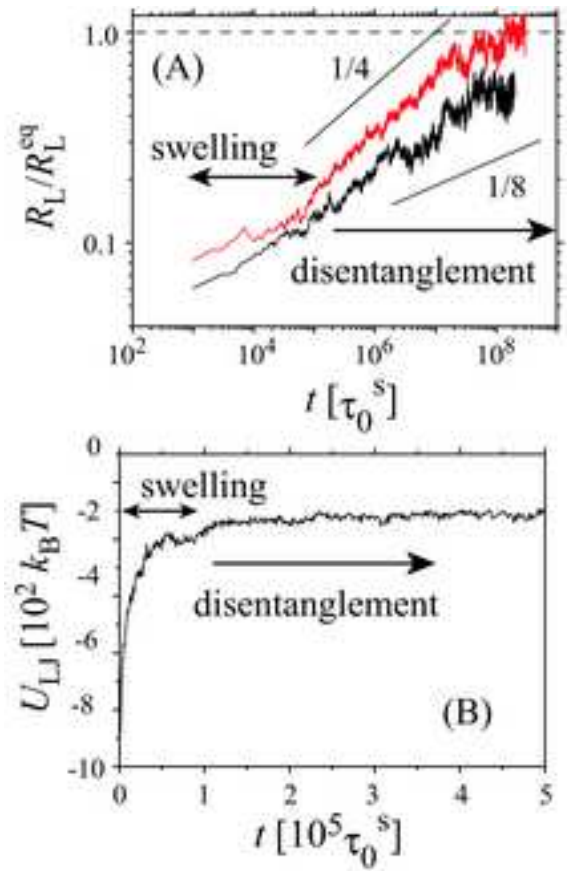}
\caption{
Time evolution of a long chain ($N=2048$) during the
 unfolding transition. (A) Time evolution  shows
 slow relaxation after initial expansion (the lower black line). The size is normalized with
 the equilibrium size, $R^{\rm eq}_{\rm L}$. \textcolor[named]{Black}{The both axes are shown
 with logarithmic scale.} The kinetics of an ideal
 polymer is also shown (the upper red line). (B) Interaction between
 monomers quickly decreases at an early stage of the transition.
 The two solid lines show $t^{1/4}$ and $t^{1/8}$.
The size at the equilibrium state is estimated
 from the additional simulations at a fixed value of epsilon.
\label{fig.long.polymer.unfolg}
}
\end{figure}

\begin{figure}
\includegraphics{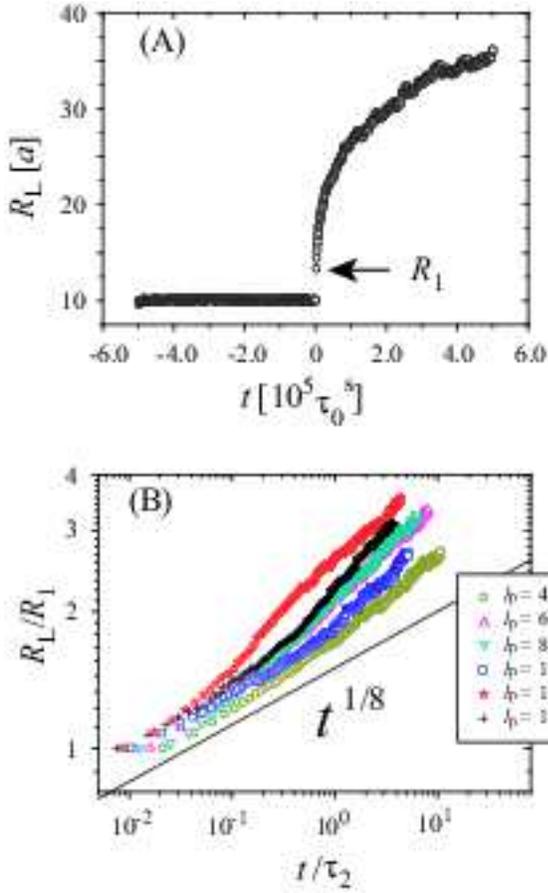}
\caption{ 
Time evolution of the long-axis length of a polymer at an
 early stage for
 various persistence length. 
Bare plot for $l_p=4$ is shown in (A).
We determine the initial values for the
 disentanglement process from $R_1$ in the figure.
\textcolor[named]{Black}{
(B) A log-log plot of the time evolution.}  Both the long-axis length and time steps
 are normalized with characteristic space and time scales (see
 Eqs. (\ref{eq.unfolding.disentanglement.scaling})). 
\label{fig:simu_unfold}
}
\end{figure}

\begin{figure}
\includegraphics{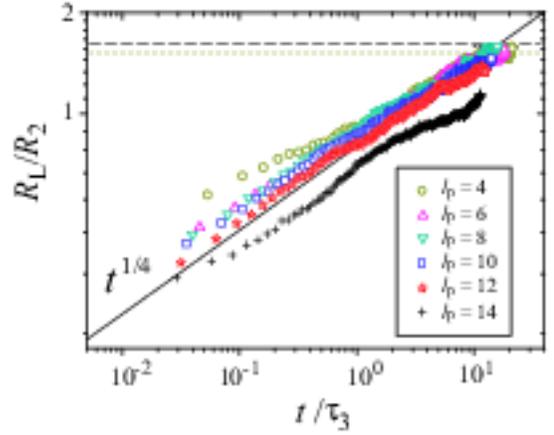}
\caption{ \textcolor[named]{Black}{
A log-log plot of time evolution of the long-axis length of a polymer at a
 later stage for
 various persistence length.} Both the long-axis length and time steps
 are normalized with characteristic space and time scales according to
 Eq. (\ref{eq.unfolding.swelling.scaling}) in the text. The dashed and
 dotted lines show the equilibrium size for $l_p=4$ and $l_p=14$, respectively.
\label{fig:simu_unfold:relaxation}
}
\end{figure}

Let us now discuss the unfolding transition.
First, we show the result of a relatively long ($N=2048$) chain.
\textcolor[named]{Black}{
Figure \ref{fig.long.polymer.unfolg} shows that there indeed exists slow
kinetics after the swelling step.
The slow kinetics is independent of the interaction energy, while the
swelling step is characterized by the interaction energy (Fig. \ref{fig.long.polymer.unfolg}(B)).
The LJ potential contains a weak attractive part even in the coil
state. 
Since the interaction is short-ranged, the energy is approximately 0 in
the coil state.
This is shown in Fig. \ref{fig.long.polymer.unfolg}B.
The system is not in equilibrium even at $10^8 \tau^{s}_{0}$, which is much
longer than the Rouse time ($\simeq 10^{6} - 10^{7}$).
On the other hand, an ideal polymer is equilibrated much faster, as expected. 
Thus, it is evident that slow kinetics exist in the unfolding
transition, and this fact is due to the non-crossing constraint of a chain with excluded
volume interaction. 
Note that the slow kinetics was also observed in the simulations of
flexible polymers in \cite{byrne:1995}.  
}
Since computation with long chains is time consuming to obtain statistics,
we used $N=256$ chains and took average over 100 runs.
Figure \ref{fig:simu_unfold} and \ref{fig:simu_unfold:relaxation} shows the results of the unfolding transition.
The size and time are normalized with characteristic space and time
scales according to Eq. (\ref{eq.theory.disentanglement.tau}) for an early stage and
Eqs. (\ref{eq.theory.relaxation.size}) and
(\ref{eq.theory.relaxation.tau}), respectively, for a later stage.
The initial size for the disentanglement step is determined from the
result of simulations.
Since the swelling process for N=256 chains is short ($\simeq 10^3$
steps), the value of $R_1$ corresponds to $R_L$ after a jump at $t > 0$ (
Fig. \ref{fig:simu_unfold}A).
The time evolution of the long-axis length exhibits a universal feature in
which the plots do not qualitatively depend on the persistence
length and the depth of quench.
In addition, the data cover from $t \simeq \tau$ to $t \gg \tau$,
and thus our simulations have a suitable time scale.  
In both figures, the slopes are in good agreement with our theory.

\section{Hydrodynamics}
\label{section:hydrodynamics}

\textcolor[named]{Black}{
With the hydrodynamic interaction our previous theory is modified depending
on their kinetics.
For expansion and contraction of a domain, it is necessary to consider the dissipation of Eq. (\ref{eq:theory:viscos}).
For cooperative motion of monomers, the modified Stokes dissipation due to hydrodynamic back
flow is applied.
}
The Stokes drag does not act on each segment, but acts on a sphere of $\xi _{\rm H}$ in size,
while hydrodynamic interactions are
screened beyond the screening length.
At a length scale smaller than $\xi _{\rm H}$, monomers move
cooperatively with the hydrodynamic back flow.
Thus, the dissipation is
\begin{equation}
 \frac{{\rm d} Q _{\rm S}}{{\rm d} t}
\simeq \eta v^2 \xi _{\rm H}.
\end{equation}
This feature is called as {\it non-drain} compared with the situation of
{\it free-drain} where
all monomers are under the Stokes drag \cite{degennes:1976,ahlrichs:2001}.

\subsection{Folding Transition}
The kinetics of the folding transition essentially does not change in the
nucleation process with hydrodynamic interaction since dissipation modifies the prefactors of the nucleation time in Eqs. (\ref{eq:fold:flexible:nucleation}) and
(\ref{folding.semiflexible.nucleationtime}).
On the other hand, in the growth process, the friction of a chain is
proportional to its size.
Therefore, the collapsed part has smaller friction, which is proportional
to $l_1^{1/3}$, for example, for flexible polymers.
The friction is much smaller than that of the coiled domain; thus,
it supports the assumption that only the collapsed domain contributes to
the dissipation in the growth step.
We obtain the velocity as 
\begin{equation}
 v = \frac{\epsilon}{\eta a^{2/3}} \frac{1}{l_1^{1/3}}
\end{equation}
for flexible polymers, and
\begin{equation}
v = \frac{\epsilon}{\eta} \left(\frac{\sigma a}{Tl_p}\right)^{2/5}
 \frac{1}{l_1^{1/5}}
\end{equation}
for semiflexible polymers.

\subsection{Unfolding Transition}
In the initial stage of the unfolding transition, the hydrodynamic
interaction is approximately screened, while in the disentanglement and the relaxation regimes, the
hydrodynamic interaction is relevant to the kinetics of the transition.
In the disentanglement process, the hydrodynamic interaction is
screened outside blobs since the blobs are pinned due to entanglements.
The dissipation inside a blob is
\begin{eqnarray}
\frac{\diff Q_{H}}{\diff t}
&\simeq& \left( \frac{L}{l_{p} g^{*}} \right)^2 
\eta R_{g} \left( \frac{\diff R_g}{\diff t} \right)^2\\
 &=& \eta \frac{L^3 l_{p}^3}{R_{L}^5}  
  \left( \frac{\diff R_{L}}{\diff t} \right)^2.
\end{eqnarray}
We obtain
\begin{eqnarray}
\left( \frac{R_{L}}{R_{disent}}\right)^{-1}
&=& 1 - t/\tau_{\rm disentangle}\\
\tau'_{\rm disentangle} &=& 
= \frac{\eta L^{5/3} l_{p}^{4/3}}{a^3}\tau_0.
\end{eqnarray}

In the relaxation regime, hydrodynamic back flow dominates the entire polymer.
Thus, the dissipation is
\begin{equation}
 \frac{{\rm d} Q _{\rm H}}{{\rm d} t}
\simeq
\eta R_L \left( \frac{{\rm d} R_L}{{\rm d} t} \right) ^2.
\label{hydrosynamics.unfold.relaxation.dissipation}
\end{equation}
By using Eqs. (\ref{theory.balance}), (\ref{eq.free.change2}), and
(\ref{hydrosynamics.unfold.relaxation.dissipation}), we obtain
\begin{eqnarray}
  \frac{R_L}{R_{\rm relax}} 
&\simeq&
\left( \frac{t}{\tau' _{\rm relax}} + 1 \right) ^{1/5},\\
\tau'_{\rm relax} &=& \eta (L l_p)^{3/2}.
\end{eqnarray}
Here, the characteristic time scale has the Zimm-type feature,
$\tau'_{\rm relax}
\sim L^{3/2}$ \cite{doi:1986}.

\section{Kinetic pathway}
\label{section:pathway}
\begin{figure}
\includegraphics{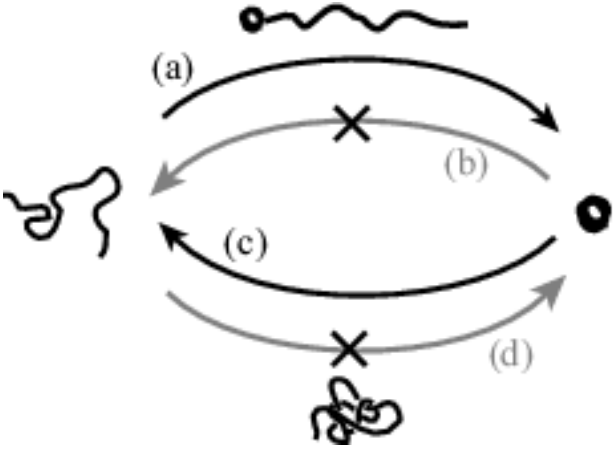}
\caption{ 
Kinetic pathways of the folding and unfolding transitions. We assume (a)
 and (c) since the pathways (b) and (d) are unrealistic, as shown in the text.
\label{fig.pathway}
}
\end{figure}

In our theory, we assume that the kinetic pathway is nucleation and growth
for the folding process (Fig.\ref{fig.pathway}a) and gradual expansion
for the unfolding process (Fig.\ref{fig.pathway}c).
However, we may also assume the opposite pathway, i.e., gradual contraction
for the folding process (Fig.\ref{fig.pathway}d) and nucleation and growth process for the
unfolding process (Fig.\ref{fig.pathway}c).
Here, we discuss the reason for why the later pathway does not appear.
An importance fact is that a faster process at the initial stage of
the transition is able to survive.
We will calculate the time evolution of the {\it unrealistic pathways}
(Fig. \ref{fig.pathway}b and d) and compare it with the results we obtained.
For simplicity, we will discuss kinetics without the hydrodynamic interaction.

\subsection{Folding transition}
Here, we consider an early stage of the kinetics of
Fig. \ref{fig.pathway}(d).
In the process, the free energy change is divided into two terms: the
elastic and the volume free energy.
The former prevents the folding transition, while the later initiates the transition.
The volume free energy is written as
\begin{equation}
F_{\rm int} \simeq 
\frac{B}{R_L^3} \left( \frac{L}{l_p} \right)^2.
\end{equation}
At $t=0$, we decrease the temperature so that $B<0$.
The free energy change is calculated as
\begin{equation}
 \frac{{\rm d} F }{{\rm d} t}
\simeq 
\left( - \frac{B L^2}{R_L^4 l_p^2} - \frac{T L l_p}{R_L^3} \right)
\frac{{\rm d} R_L}{{\rm d} t},
\label{pathway.folding.freeenergychange}
\end{equation}
where the second term is the contribution of the elastic free energy.
The first term contributes to decrease the free energy, while the
second makes a opposite contribution.
When $R_L$ is close to the equilibrium size in the swollen state $\simeq
(Ll_p)^{1/2}$, the first term is proportional to $\sim L^0$ and the
second term is proportional to $\sim L^{-1/2}$.
We are interested in the kinetics of the early stage of this pathway.
Thus, we neglect the second term in (\ref{pathway.folding.freeenergychange}).
The dissipation in this process is
\begin{equation}
 \frac{{\rm d} Q _{\rm S}}{{\rm d} t}
\simeq \eta L \left( \frac{{\rm d} R_L}{{\rm d} t} \right) ^2,
\label{entropy_swell}
\end{equation}
where we neglect logarithmic factor for simplicity.
With Eq. (\ref{theory.balance}), the time evolution is
\begin{equation}
\frac{R_L}{R_2} \simeq 
\left( 1 - \frac{|B|L}{l_p^2 R_2^5}t \right)^{1/5}.
\end{equation}
At the early stage, the time evolution is 
\begin{equation}
 R_L \simeq R_2 - \frac{|B|L}{l_p^2 R_2^4}t,
\end{equation}
where the velocity is proportional to $L^{-1}$. 
Thus, we expect this process to be too slow to proceed before the nucleation process.

\subsection{Unfolding transition}
In the unfolding transition, we may assume the kinetic pathway of Fig.~\ref{fig.pathway}(b), that is, the
collapsed polymer with a short unfolded coiled part.
The kinetics is driven by the free energy change of the peeled part,
which is in fact the difference of the free energy
between the coiled and the collapsed states:
\begin{equation}
 \frac{\diff F}{\diff t} \simeq 
-\frac{T}{l^{2/3} l_p^{1/3}} \frac{\diff l}{\diff t}.
\end{equation}
The dissipation from the coiled part is
\begin{equation}
 \frac{\diff Q}{\diff t} \simeq \eta l \left( \frac{\diff l}{\diff t} \right)^2.
\end{equation}
Therefore, we obtain 
\begin{equation}
 l \simeq \left( \frac{T^3}{\eta^3 l_p} \right)^{1/8} t^{3/8},
\end{equation}
which indicates that the characteristic time for this pathway is proportional to $L^{8/3}$.
As in section~\ref{section:unfolding}, the swelling regime proceeds much
faster.
From this, we conclude that the pathway is not realized.

\section{Summary and Remarks}
\label{section:summary}

In summary, we investigated the kinetics of the folding and unfolding
transitions in a single semiflexible polymer with and without the
hydrodynamic interaction.
We found that the velocity of the length of the collapsed domain depends
inversely on the length in the folding transition, and the dynamic scaling
exponents are $1/8$ and $1/4$ for the disentanglement and relaxation
steps, respectively, in the unfolding transition without the
hydrodynamic interaction.
The time dependence without the hydrodynamic interaction is also
calculated using Langevin Dynamics simulations, is found to be in good
agreement with our theory.

We discussed the origin of dissipation during the folding and unfolding
transitions.
The main contribution in the folding transition is found to be the motion of
the collapsed domain along a chain.
We proposed a slow relaxation regime in the unfolding transition arising
from entanglements of a polymer.
Since this regime is not found in an ideal polymer, we consider that the slow
kinetics originate from the topological constraint due to the excluded
volume interaction.

Although the kinetics of single semiflexible polymers have not been studied
extensively in experiments, the authors in ref. \cite{yoshikawa:1996b,ichikawa:2004}
investigated the time evolution of long DNA molecules by using fluorescent microscopy.
In \cite{ichikawa:2004}, DNA molecules were contracted using optical tweezers under good
solvent conditions.
After switching the optical tweezers off, the time evolution
of the size of DNA molecules was observed, and small exponent of 0.125
was found.
This result corresponds well with the disentanglement regime in
our results.
In \cite{yoshikawa:1996b}, the transitions were induced by a sudden change
in the concentration of multivalent cations.
In the folding transition, they found linear time dependence of the
apparent size of a DNA molecule.
Since the data have large fluctuations, quantitative comparison is
difficult at the moment.
Nevertheless, we believe that further experimental studies would be helpful
for comparison between theory and experiments.

To conclude this paper, we make some remarks for future investigations.
\begin{enumerate}
 \item We have considered single nucleation along a chain.
This is a limited situation because nucleation usually occurs
       simultaneously along a chain. 
However, as we discussed, the nucleation time depends exponentially on
       the persistence length.
This indicates that we need a very long chain to obtain multiple
       nucleations with a large persistence length.
This is in contrast to the case of flexible polymers where multiple
       nucleations called {\it pearl-necklace structure} dominates at an
       early stage of the folding transition \cite{klushin:1998,halperin:2000}.
We should note that when we have a very long semiflexible chain, we
       should consider a network structure rather than multiple nucleations
       along a chain since nucleation does not occur locally due to the
       bending rigidity.

\item As noted in introduction, we have concentrated on the toroidal
       structure.
This makes the problem tractable.
To discuss further, we have to consider cylindrical conformations.
They appear as the equilibrium state in a range of parameters, but the
       difficulty is that they have many metastable structures
       depending on the number of times they are folded.
Cylinders that are few times folded are likely to appear in the kinetics though
       they are not equilibrium structure but metastable states.
After
       certain time, they become more folded structures or sometimes make a transition into
       toroidal structures.
The kinetics of folding into cylindrical conformations are
       dominated by such hopping steps, which are different from the
       kinetics for toroidal conformations.
It is interesting, as a further study, to discuss the
       bifurcation between the two kinetics, the nucleation and growth,
       which are discussed in the present work, and hopping steps found
       in kinetics of cylindrical collapses.     

\item Our calculation of the nucleation time is qualitative. 
       In order to make quantitative discussions, it is necessary to take loop
       formation into account. 
       In fact, the nucleation process initiates with loop formation
       \cite{hyeon:2006}.
       This modifies the prefactors in the nucleation time
       Eqs. (\ref{eq:fold:flexible:nucleation}) and
       (\ref{folding.semiflexible.nucleationtime}).
       Nevertheless, the nucleation time is dominated by the exponential
       factor. Therefore, we expect our results to be qualitatively reasonable.

\item Our comparison between theory and simulation is under the
       condition without the hydrodynamic interaction.
       Recently, several algorithms have become available to deal with the hydrodynamic interaction in
       polymer systems.
       For example, Stochastic Rotation Dynamics
       \cite{kikuchi:2002,kikuchi:2005} and Lattice Boltzmann \cite{duenweg:2004} were
       implemented for investigating the kinetics of polymers.
       Such simulations would reveal details of the physical processes
       involved in
       the kinetics of the folding and unfolding transitions when the
       hydrodynamic interaction is also considered, and would test the
       validity of our theory.

\item Quantitative arguments for the disentanglement process are still
       left as a future study.
In particular, the crossover between disentangled and swollen states is not
       clear.
We should note that the characteristic
       time in this study is overestimated.
In fact, disentanglement process does not end at $g^{*} \simeq L/l_p$,
       but is replaced earlier by the relaxation process since topological blobs suddenly disappear when they
       become sufficiently large.
Our theory does not include such effects.
\end{enumerate}

\begin{acknowledgments}
The author is grateful to T. Sakaue, Y. Murayama, K. Yoshikawa and T. Ohta for helpful discussions.
We would like to acknowledge the support of a fellowship from the JSPS
 (No. 7662).
Part of the numerical calculations in this work was carried out on
 Altix3700 BX2 at YITP in Kyoto University.
\end{acknowledgments}


\begin{thebibliography}{36}
\bibitem[{not()}]{note1}
\bibinfo{note}{Since we discuss the spontaneous transition, we need not to
  consider the external work. In the case of single polymers under mechanical
  force, the external work ${\rm d} W / {\rm d} t$ have to be
  considered.\cite{yoshinaga:2005}}.
\end{thebibliography}

\end{document}